\documentclass[aps,prb,twocolumn,showpacs,citeautoscript,reprint,longbibliography]{revtex4-2}
\usepackage{boldline,multirow}
\usepackage{float}
\usepackage{graphicx,natbib}
\usepackage{bm,amsmath,amssymb,wasysym,amsfonts,dcolumn}
\usepackage[colorlinks=true, urlcolor=blue, linkcolor=blue, citecolor=blue, pdfborder={0 0 0}]{hyperref}
\usepackage{ulem}
\usepackage[usenames,dvipsnames]{xcolor}
\usepackage{lipsum}
\usepackage{epstopdf}
\usepackage{cleveref}
\crefname{figure}{Fig.}{Figs}
\crefname{table}{Table}{Tables}
\crefname{section}{Sec.}{Sections}

\newcommand{\ket}[1]{\left| #1 \right>} % for Dirac bras
\newcommand{\bra}[1]{\left< #1 \right|} % for Dirac kets
\newcommand{\braket}[2]{\left< #1 \vphantom{#2} \right|
 \left. #2 \vphantom{#1} \right>} % for Dirac brackets
\begin{document}

%%%%%%%%10%%%%%%%%20%%%%%%%%30%%%%%%%%40%%%%%%%%50%%%%%%%%60%%%%%%%%70%%%%%%%%80
\title{Orbital Hall effect in transition metals from first-principles scattering calculations}
		
\author{Max Rang}
\author{Paul J. Kelly}
\affiliation{Faculty of Science and Technology and MESA$^+$ Institute for Nanotechnology, University of Twente, P.O. Box 217,
	7500 AE Enschede, The Netherlands}

\date{\today}

\begin{abstract}
We use first-principles scattering calculations based upon wave-function matching and implemented with a tight-binding MTO basis to evaluate the orbital Hall conductivity $\sigma_{\rm oH}$ for Ti, V, Cr, Cu and Pt metals with temperature-induced lattice disorder. 
Only interatomic fluxes of orbital angular momentum are included in these estimates; intraatomic fluxes which do not contribute to the transfer of angular momentum are explicitly excluded.
The resistivity and orbital Hall angle are both found to be linear in temperature so $\sigma_{\rm oH}$ is at most weakly temperature dependent. 
The value of $\sigma_{\rm oH}$ we obtain for bulk Cr is $ \approx 2 \times 10^3
(\hbar/e) \, (\Omega \, {\rm cm})^{-1}$ which is substantially lower than previously obtained theoretical results (but in better agreement with values extracted from experiment). In units of $10^3 (\hbar/e) (\Omega \, {\rm cm})^{-1}$, the values of $\sigma_{\rm oH}$ obtained for Ti, V and Pt are $5$, $6$ and $7$, respectively.
\end{abstract}


\maketitle

\section{Introduction}
The orbital Hall effect (OHE) \cite{Bernevig:prl05, Tanaka:prb08, Kontani:prl09}, that is the orbital analogue of the spin Hall effect (SHE) \cite{Dyakonov:zetf71, Dyakonov:pla71, Hirsch:prl99, Hoffmann:ieeem13, Sinova:rmp15}, has recently attracted a lot of attention as a source of a current of angular momentum.
It is considered to be more fundamental than the SHE \cite{Go:prl18} and predicted to be significantly larger in size, in particular in systems with weak spin-orbit coupling (SOC) \cite{Jo:prb18, Salemi:prm22}.
Its magnitude is typically estimated using a Kubo-like expression for the orbital angular momentum (OAM) conductivity that is analogous to that used to determine the conventional charge conductivity; the similarity of this expression to that for the anomalous Hall effect (AHE) has led to it being called the ``orbital Berry curvature'' \cite{Baek:prb21, Bhowal:prb21}.
Computation of the orbital Hall conductivity (OHC) then involves the integration of correlation functions between the OAM current operator ${\bf j}_l$ with elements $j^{\beta}_{l\gamma} = \frac{1}{2}\lbrace \upsilon_{\beta}, l_{\gamma} \rbrace$ and the current operator ${\bm \upsilon}$ with elements $\upsilon_{\alpha}$ where $\lbrace A, B \rbrace$ is the anticommutator and $\alpha, \beta, \gamma \in \lbrace x,y,z \rbrace$ \cite{footnote2}; a driving current of charge in the $\alpha$ direction induces a current of OAM in the $\beta$ direction that is polarized in the orthogonal $\gamma$ direction. 

In a basis of Wannier functions $\phi_R$ centred on ${\bf R}$, the angular momentum can be decomposed into a conventional ``intra-atomic'' term  $\langle \phi_R | {({\bf r-R}) \times {\bm \upsilon}} | \phi_R \big\rangle$ and an additional ``extra-atomic'' term ${\bf R} \times \langle \phi_R | {\bm \upsilon} | \phi_R \rangle$ that has been shown to be comparable in magnitude to the intra-atomic term in spite of $\langle \phi_R | {\bm \upsilon} | \phi_R \rangle$ vanishing for bulk Wannier functions \cite{Thonhauser:prl05}. 
In calculations of the OHE, this term is more often than not neglected \cite{Tanaka:prb08, Kontani:prl09, Go:prr20b, Salemi:prm22} although in the Kubo formalism for infinite ``bulk'' systems, it is predicted to be substantial for materials like gapped graphene \cite{Bhowal:prb21}, MoS$_2$ bilayers \cite{Cysne:prb22} and in the narrow bandgap semiconductors SnTe and PbTe as well as the transition metals V and Pt \cite{Pezo:prb22}. 

Currents are conventionally understood as the flux of some property through a volume and the meaning of the current of a property like ${\bf R} \times \langle \phi_R | {\bm \upsilon} | \phi_R \rangle$ is unclear. 
Indeed, to define a current one must define some volume, which necessarily localizes the corresponding property to that volume.
Our intuitive understanding of the OHE, as of the AHE and SHE, involves electrons hopping interatomically acquiring an anomalous velocity (owing to the Berry curvature). 
This current is perpendicular to the charge current in general but, depending on the crystal symmetry, can be in any direction \cite{Seemann:prb15, Roy:prm22}.

The Kubo expression that is conventionally \cite{Jo:prb18, Salemi:prm22} employed to calculate the ``$X$'' conductivity tensor, where $X$ is charge ($c$), spin angular momentum ($s$) or OAM ($l$), is
\begin{equation}
\label{Eq:K1}
 \sigma^{X\gamma}_{\alpha \beta} = \frac{e}{\hbar} \sum_n \int \frac{d^3{\bf k}}{\left(2\pi\right)^3} f_{n\mathbf{k}} \Omega^{X\gamma}_{n,\alpha \beta}(\mathbf{k}),\\
\end{equation}
 where
 \begin{equation}
\label{Eq:K2}
\Omega^{X{\gamma}}_{n,\alpha \beta}({\bf k}) = 2\hbar^2 \!\! \sum_{m\ne n} \! {\rm Im} \! 
\left[  \frac{\bra{u_{n{\bf k}}} j^{\beta}_{X{\gamma}} \ket{u_{m{\bf k}}}    
        \bra{u_{m{\bf k}}}v_{\alpha}\ket{u_{n{\bf k}}}}  
        {\left(\varepsilon_{n{\bf k}} - \varepsilon_{m{\bf k}}\right)^2} \right].
\end{equation}
$f_{n\mathbf{k}}$ is the Fermi-Dirac distribution, $\ket{u_{n\mathbf{k}}}$ is the cell-periodic part of the Bloch state with energy eigenvalue $\varepsilon_{n\mathbf{k}}$ and the velocity operator $\upsilon_\alpha$ is usually chosen to be  
\begin{equation}
\upsilon_{\alpha} = \frac{1}{\hbar}\frac{\partial H({\bf k})}{\partial k_\alpha}
\end{equation}
in the crystal momentum representation \cite{Callaway:74}. 
Though conceptually simple, this $\mathbf{k}$-space formulation of the velocity is beset with mathematical nuance \cite{Esteve-Paredes:sp23} and must be done with care.
Evaluation of \eqref{Eq:K1} and \eqref{Eq:K2} requires a very fine discretization of the reciprocal space combined with Wannier interpolation onto the corresponding dense mesh \cite{Zeer:prm22, Go:prb24}. 
Performing the calculations in the primitive unit cell of the periodic crystal precludes the direct simulation of disorder.

In this work, we address the orbital Hall conductivity from the point of view of scattering calculations.
First, rather than defining the OAM current operator as the orbital analogue of the conventional spin current operator, we begin with a continuity equation for the OAM density on each atom and define the current as the flux of OAM between atoms. 
Experimentally, one is interested in the transfer of OAM from some source layer of nonmagnetic material into an adjacent layer of ferromagnetic material. 
For this to be measurable, it must happen on a length scale that is at least as large as the interatomic distance. 
By deriving the OAM current from an interatomic flux, we satisfy this requirement theoretically. 
The precise mathematical relation between the Kubo formula and the method we propose here poses an interesting theoretical challenge, especially as it pertains to currents of spin and orbital angular momenta, but is beyond the scope of the present paper. 
We will see that the OAM current operator we derive is closely related to the definition of the spin current operator advocated by Niu {\it et al.} \cite{Shi:prl06, *ZhangP:prb08} that is still however the topic of discussion \cite{Marcelli:ahp21}.

Second, the contribution of atomic disorder, be it chemical (e.g. impurities) or structural (arising from thermal disorder, stacking faults, vacancies, self-interstitials etc.) is not addressed in standard Kubo formula approaches \cite{Jo:prb18, Salemi:prm22, Pezo:prb22, Cysne:prb22, Cysne:prl21, Bhowal:prb20a, Bhowal:prb20b, Go:prl18}. 
Such disorder is present in all samples used in real experiments and its role in determining the OHC needs to be understood.
To address this second issue, we will perform quantum mechanical scattering calculations using a wave-function matching (WFM) method \cite{Ando:prb91} implemented \cite{Xia:prb06, Starikov:prb18, Wesselink:prb19} in a basis of tight-binding (TB) muffin-tin orbitals (MTO) \cite{Andersen:prl84, *Andersen:85, *Andersen:prb86}. 
This method allows us to decompose spin and charge currents in terms of the bond currents between atoms. Indeed, in the scattering formalism, only the ${\bm \upsilon}_{\rm inter}$ term contributes to long-range transport. 
The Twente Quantum Transport ({\sc tqt}) code \cite{TQT:GitHub} has been extended to allow the computation of orbital currents. The relatively low computational cost of running these calculations allows for the treatment of large supercells so that disorder can be treated in real space \cite{LiuY:prb11, *LiuY:prb15}. In practice, the upper limit for the system size is so large that realistic interfaces, bilayers and thin films can be studied \cite{Gupta:prl20, Nair:prb21a, Nair:prb21b, Gupta:prb21, LiuRX:prb22, Gupta:prb22}.
The formalism of bond currents in the atomic spheres approximation allows us to use the conventional OAM operator while still appropriately treating the current on an interatomic footing.
To wit, the definition of the velocity operator in the scattering calculations is
\begin{equation}
\label{eq:veloc}
{\bm \upsilon} = \frac{1}{i\hbar} \left[ \mathbf{r}, H\right],
\end{equation}
avoiding problems associated with the $\mathbf{k}$-space definition of the velocity \cite{Esteve-Paredes:sp23}.

\section{Method}
\label{Sec:Method}
A brief overview of the {\sc tqt} code \cite{TQT:GitHub} used in this study is given in \Cref{ss:wfm}.
The extension of the WFM formalism implemented in {\sc tqt} to calculate currents of OAM is presented in \Cref{ss:oc}. 
The treatment of disorder as well as additional computational details are described in \Cref{ss:discomp}.

\subsection{Wave function matching}
\label{ss:wfm}
\begin{figure}[t]
\includegraphics[width=8.6cm]{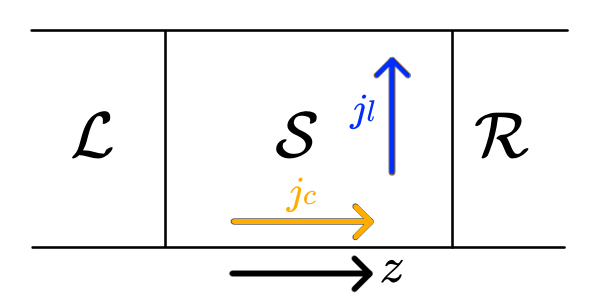}
\caption{Schematic of the scattering geometry. The charge current $j_c$ is parallel to the $z$ direction and the orbital current $j_l$ perpendicular to it.
}
\label{fig:schematic}
\end{figure}

The scattering geometry consists of three regions, a left lead ($\mathcal{L}$), a right lead ($\mathcal{R}$) and a scattering region ($\mathcal{S}$);  see \Cref{fig:schematic}. 
Left and right leads that are infinitely long in the $-z$ and $z$ directions, respectively,  sandwich the scattering region in the middle.

%%%%%%%%%%%%%%%%%%%%%%%%%%%% Fig.1

The time-independent Schr\"{o}dinger equation is solved for the scattering region with the open boundary conditions corresponding to semi-infinite leads represented by a ``self-energy'' $\hat{\Sigma}$, an energy-dependent potential in the lead layers next to the scattering region. 
The set of equations to be solved are the system of linear equations
\begin{equation}
\left(\varepsilon \hat{I} - \hat{H} - \hat{\Sigma}\right)\Psi = \hat{Q},
\end{equation}
where $\varepsilon$ is the energy of the incoming wave (usually the Fermi energy), $\hat{H}$ is the Hamiltonian of the scattering region, $\hat{\Sigma}$ is the self-energy of the coupling with the lead states, $\Psi$ is the wave function and $\hat{Q}$ is a source term. 
The latter are incoming Bloch states, either from the left lead or from the right lead. 
In the $z$ direction, the geometry is infinitely long (or finitely long, but with open boundary conditions, according to one's taste). In the $x$ and $y$ directions, we use periodic boundary conditions and ``lateral'' supercells to simulate disordered materials. 
In practice, when modelling thermal disorder \cite{LiuY:prb11, *LiuY:prb15}, we find that transport properties scarcely change when the supercell size is increased systematically in the $x$ and $y$ directions even when the supercells are as small as 5$\times$5 \cite{Starikov:prb18, Wesselink:prb19}.
Further mathematical and computational details of the method can be found in Refs \cite{Starikov:prb18, Khomyakov:prb05}.

\subsection{Orbital currents}
\label{ss:oc}

Starting with the continuity equation, we derive an expression for the flux of OAM between atoms and discuss the limitations of the approach. 
Since the wave function is expanded in a localized orbital (TB-MTO) basis, we can write
\begin{equation}
\ket{\Psi} = \sum_{P} \hat{P} \ket{\Psi} = \sum_{P} \ket{\Psi_P},
\end{equation}
where $P$ runs over the atoms in the scattering region \cite{Wesselink:prb19, Nair:prb21a}. 
In the atomic spheres approximation (ASA), the basis functions centered on different sites $P$ and $Q$ are (almost) orthogonal, i.e., $\braket{\Psi_P}{\Psi_Q} \approx \delta_{PQ}$ \cite{footnote3} and we can decompose any local observable as
\begin{equation}
\bra{\Psi}\hat{A}\ket{\Psi} = \sum_{P} \bra{\Psi_P}\hat{A}\ket{\Psi_P}.
\end{equation}
Assuming that $\hat{\ell}$ is local, i.e., $\bra{\Psi_P} \hat{\ell}\ket{\Psi_Q} \propto \delta_{PQ}$, allows us to compute the time derivative of the expectation value of the OAM. Use of the continuity equation relates this to the flux of OAM out of or into atom $P$
%\begin{equation}
%\label{eq:cont}
%\begin{split}
%\frac{d}{dt}\bra{\Psi_P}\hat{\ell}\ket{\Psi_P} &= \bra{\frac{d\Psi_P}{dt}} \hat{\ell} \ket{\Psi_P} + \bra{\Psi_P}\hat{\ell} \ket{\frac{d\Psi_P}{dt}}\\
%&= \frac{1}{i\hbar} \left[ \bra{\Psi_P} \hat{\ell} \hat{H} \ket{\Psi} - \bra{\Psi}\hat{H}\hat{\ell}\ket{\Psi_P}\right]\\
%&= \sum_{Q} j^{PQ}_{\hat{\ell}}.
%\end{split}
%\end{equation}
%PJK
\begin{equation}
\begin{split}
\frac{d}{dt}\bra{\Psi_P}\hat{\ell}\ket{\Psi_P} = \bra{\frac{d\Psi_P}{dt}} \hat{\ell} \ket{\Psi_P} + \bra{\Psi_P}\hat{\ell} \ket{\frac{d\Psi_P}{dt}}\\
= \frac{1}{i\hbar} \left[ \bra{\Psi_P} \hat{\ell} \hat{H} \ket{\Psi} - \bra{\Psi}\hat{H}\hat{\ell}\ket{\Psi_P}\right] = \sum_{Q} j^{PQ}_{\hat{\ell}}.
\end{split}
\end{equation}
%PJK
This equation is completely analogous to a derivation of spin currents in the WFM TB-MTO method and more details can be found in Ref.~\cite{Wesselink:prb19}. 
The interatomic flux of OAM polarized in the $\alpha$ direction can be written
\begin{equation}
\label{eq:jlpq}
\!\! j^{PQ}_{\hat{\ell}_\alpha} = \frac{1}{i\hbar} \left[ \bra{\Psi_P} \hat{\ell}_\alpha \hat{H}_{PQ} \ket{\Psi_Q} - \bra{\Psi_Q} \hat{H}_{QP} \hat{\ell}_{\alpha} \ket{\Psi_P} \right],
\end{equation}
and, like the spin, the orbital angular momentum is not a conserved quantity so $j^{PQ}_{\hat{\ell}_\alpha} \ne -j^{QP}_{\hat{\ell}_\alpha}$. 
To transform the expression \eqref{eq:jlpq} from a flux into a current, it needs to be multiplied by $\mathbf{d}_{PQ} = \mathbf{r}_P - \mathbf{r}_Q$, where $\mathbf{r}_i$ is the position of atom $i$. 
Then, the current of OAM between $P$ and $Q$ is 
\begin{equation}
\label{eq:curPQ}
\begin{split}
\bar{j}^{PQ}_{\hat{\ell}_\alpha} &= \frac{1}{2}\left(\mathbf{r}_P -\mathbf{r}_Q \right) j^{PQ}_{\hat{\ell}_\alpha} + \frac{1}{2}\left(\mathbf{r}_Q -\mathbf{r}_P \right) j^{QP}_{\hat{\ell}_\alpha} \\
&=\frac{\mathbf{r}_P - \mathbf{r}_Q} {2i\hbar} \left[ \bra{\Psi_P} \hat{\ell}_\alpha \hat{H}_{PQ} + \hat{H}_{PQ}\hat{\ell}_\alpha \ket{\Psi_Q} \right. \\
&\phantom{aaaaaaaa.}\left. - \bra{\Psi_Q} \hat{\ell}_\alpha \hat{H}_{QP} + \hat{H}_{QP} \hat{\ell}_\alpha \ket{\Psi_P} \right].
\end{split}
\end{equation}
This is the expression that is used to obtain the results presented in \Cref{Sec:Results} (if the atoms $P$ and $Q$ are in the same cell that the orbital current is being averaged over; if they are in different cells, the expression becomes slightly more involved \cite{Wesselink:prb19}).
In practice, we use \eqref{eq:jlpq} to calculate interatomic fluxes which are then processed by analogy with their spin-current-density tensor counterparts $\tensor{j_s}$ to calculate orbital-current-density tensors $\tensor{j_{\ell}}$ that are either layer-averaged, $\tensor{j_{\ell}}(z)$ \cite{Wesselink:prb19}, or fully spatially resolved,  $\tensor{j_{\ell}}({\bf r})$ \cite{Nair:prb21a}.

\subsection{Disorder and some computational details}
\label{ss:discomp}
When implemented with a minimal basis of localized TB-MTOs \cite{Andersen:prl84, *Andersen:85, *Andersen:prb86}, the wave-function matching formalism \cite{Ando:prb91} results in a highly sparse system of linear equations \cite{Xia:prb06, Starikov:prb18} that we solve using the {\sc mumps} package \cite{Amestoy:cmame00, Amestoy:siamjm01, Amestoy:pc06}.  
The highly efficient sparse solver makes it possible to routinely handle scattering regions containing $10^4-10^5$ atoms but these cannot be distributed arbitrarily in each dimension because of how the memory requirements scale with the lateral supercell size in the $x$ and $y$ directions. This constraint is related to the boundary condition imposed by the Bloch nature of the lead states in the ``embedding layer'', the last lead layer at the interface between a lead and the scattering region. 
In a localized orbital representation, the Hamiltonian of the scattering region is highly diagonal. 
However, the Bloch eigenstates of the leads have a finite amplitude for these localized orbitals on all atoms of the embedding layer so that the Hamiltonian becomes dense; matrix operations involving the embedding layer ultimately limit the size of problems that can be addressed. 
As a result, the limit of the lateral supercell size we can currently treat is roughly 20$\times$20 for close-packed systems like Pt with a minimal $spd$ orbital basis when spin-orbit coupling is included. 
Still, this supercell is sufficiently large to allow us to construct realistic geometries for systems of current experimental interest \cite{Stamm:prl17} like Pt thin films \cite{Nair:prb21b}.

Thermal lattice disorder is modelled by randomly displacing atoms from their equilibrium positions according to a Gaussian distribution. 
The variance of this Gaussian is chosen to reproduce the experimental resistivity at a particular temperature \cite{Wesselink:prb19, Nair:prl21}. 
Acceptably converged results are obtained using 5-10 randomly disordered configurations over which transport properties are averaged. 

Determining transport properties requires a summation over all propagating states at the Fermi energy in the leads. 
This requires sampling the two-dimensional (2D) Brillouin zone (BZ) corresponding to the in-plane translational symmetry of the scattering region. 
In practice, we sample the 2D BZ uniformly and experience has shown that a 160$\times$160 grid produces very well converged results for a 1$\times$1 unit cell \cite{Xia:prb06, Starikov:prb18, Wesselink:prb19}. 
When a lateral supercell is used, the BZ is folded down so that multiple ${\bf k}$-points of the original 1$\times$1 unit cell can be mapped onto a single ${\bf k}$-point, which means that to obtain the same accuracy, fewer ${\bf k}$-points are needed \cite{Xia:prb06}. 
From now on, we will specify the 1$\times$1 equivalent number of $\mathbf{k}$-points. 
For example, a 160$\times$160 sampling of the first Brillouin zone (BZ) for a 1$\times$1 real space system is equivalent to a 16$\times$16 sampling of the BZ for a 10$\times$10 lateral supercell. 
The numbers will not always be nice and round because the $\Gamma$ point is included and the number of ${\bf k}$-points is hence odd.
In this manuscript, all results were obtained using a $7\times7$ lateral supercell. 
To gain additional insight, we have performed calculations where the Fermi energy is varied and the transport calculation is repeated for the same configuration multiple times, each time at a different energy.

\section{Results}
\label{Sec:Results}
We proceed to calculate the orbital Hall conductivity (OHC) and spin Hall conductivity (SHC) for a number of selected transition metals. 
In spite of their weak spin-orbit interaction, there is a lot of interest in the 3$d$ transition metals bcc Cr and V that are reported to exhibit large orbital Hall angles \cite{Jo:prb18, Sala:prr22} as well as for the prototypical and much-used spin Hall material, fcc Pt \cite{Kimura:prl07, Guo:prl08, Sinova:rmp15}.
To obtain Kohn-Sham ASA potentials for bulk materials, DFT calculations are performed for  perfect crystals using the {\sc questaal} code \cite{Pashov:cpc20}. 
Specifically, the von Barth-Hedin exchange-correlation functional was used \cite{vonBarth:jpc72}, with spin-orbit coupling included self-consistently because of the importance of determining the Fermi energy accurately \cite{Wesselink:prb19}.
A minimal $spd$ basis of TB-MTOs was used since the transport calculations will be performed using that same basis. 
As a convergence check, an $spdf$ basis was additionally used for Cr. 
The self-consistent DFT calculations were performed on an 18$\times 18 \times$18 ${\bf k}$-point grid.

The orbital Hall effect can be characterized quantitatively in two different ways. 
The orbital Hall angle is the ratio of the perpendicular orbital Hall current to the driving charge current which becomes dimensionless when currents of charge and angular momentum are expressed as particle currents \cite{footnote2}. 
Alternatively, one can compute the orbital Hall conductivity, expressed in units of $\hbar/e \left(\Omega \, {\rm cm}\right)^{-1}$.   
The linear response transport formalism we use lends itself naturally to the former since all response currents are expressed per unit driving current. 
However, in order to directly compare to literature results, we report conductivities here.

Resistivities can be determined by calculating the length dependence of the configuration-averaged resistance \cite{Starikov:prb18, Wesselink:prb19, Nair:prl21} requiring calculations for of order five different lengths.
Alternatively, one can compute the chemical potential as a function of the position in the scattering region in the direction of transport and obtain the resistivity from a single configuration \cite{Wesselink:unpublished14}. The key idea is to determine the slope of the chemical potential within the scattering region by fitting.
We can compute the orbital Hall conductivity at some band filling by dividing the average orbital Hall angle within the scattering region by the resistivity computed at that band filling, $\sigma_{\rm oH} = \Theta_{\rm oH}/{\rho}$, using the fact that the resistivity $\rho$ is the inverse of the (charge) conductivity $\sigma_c$ and using the definition of the orbital Hall angle $\Theta_{\rm oH} = j_{\rm oH}/j_c = \sigma_{\rm oH} E / \sigma_c E$, where $E$ is the electric field.

\subsection{bcc Cr}
\subsubsection{Band filling}

%%%%%%%%%%%%%%%%%%%%%%%%%%%% Fig.2
\begin{figure}[t]
\includegraphics[width=8.6cm]{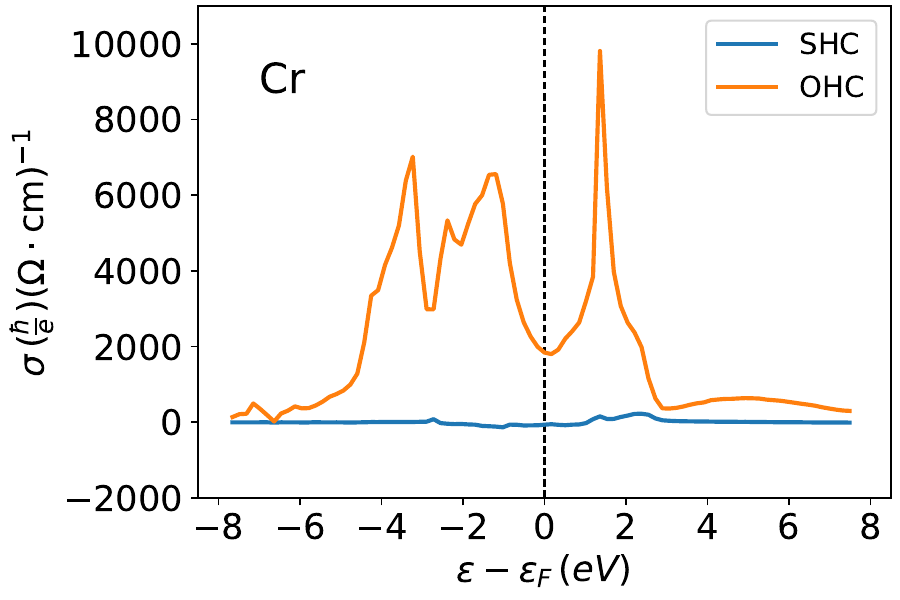}
\caption{Energy dependence of the orbital- and spin-Hall conductivities for bulk Cr with thermal lattice disorder corresponding to $T=300 \,$K.
The displacements are drawn from a random distribution with a root mean square (rms) displacement of $\Delta = 0.085 \AA$. 
}\label{fig:eF_OHC_Cr}
\end{figure}

%%%%%%%%%%%%%%%%%%%%%%%%%%%% Fig.3
\begin{figure}[b]
\includegraphics[width=8.6cm]{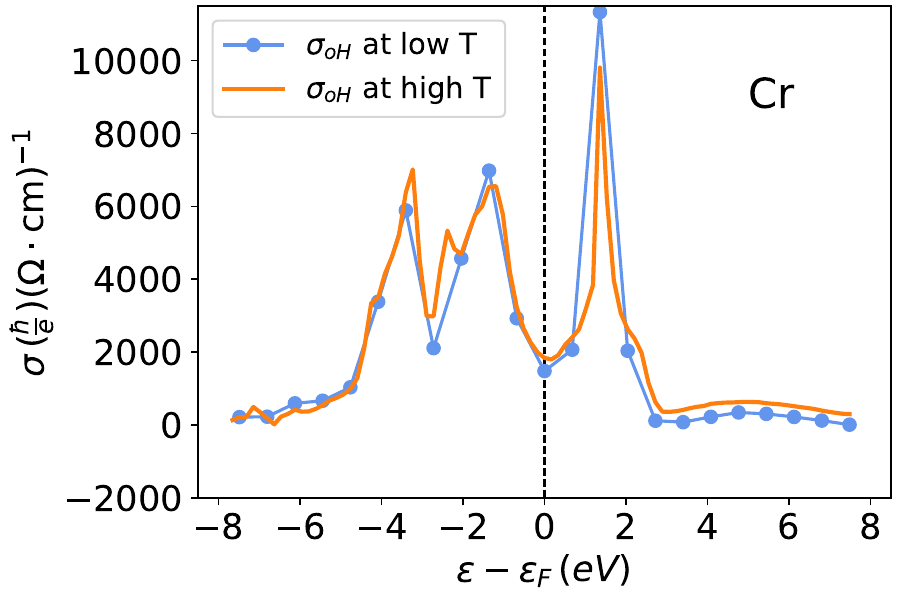}
\caption{Energy dependence of the orbital Hall conductivity for Cr for two different values of the rms displacements used to generate thermal lattice disorder. Two different variances are chosen for the displacements of the atoms leading to a ``high temperature'' and a ``low temperature'' geometry. These values are $\Delta_{\text{high}} = 0.085 \AA$ and $\Delta_{\text{low}} = 0.032 \AA$.
}
\label{fig:eF_OHC_Cr_lowT}
\end{figure}

The dependence of the spin-Hall and orbital-Hall conductivities calculated for a single configuration of thermally disordered bcc Cr is shown in \Cref{fig:eF_OHC_Cr} as a function of the energy in a rigid band approximation. 
At the Fermi energy $\varepsilon_{\rm F}$, the value of $\sigma_{\rm oH}$ is $ \approx 2 \times 10^3 (\hbar/e) (\Omega \, {\rm cm})^{-1}$; the small SOC for 3$d$ metals leads to a value of $\sigma_{\rm sH}$ that is so small it is virtually indistinguishable from the noise inherent in a single-shot transport calculation, i.e. $\sim 10 (\hbar/e) (\Omega\, {\rm cm})^{-1}$. 
Taking a different realisation of disorder with the same variance changes $\sigma_{\rm oH}(\varepsilon)$ only marginally. 
In fact, taking a smaller value for the variance of the displacements, corresponding to a lower temperature, only alters the orbital Hall conductivities obtained this way slightly.
The low and high temperatures for which the OHC is shown in \Cref{fig:eF_OHC_Cr_lowT} correspond to resistivities $\rho \approx 2 \mu\Omega \, {\rm cm}$ and $\rho \approx 11 \mu\Omega \, {\rm cm}$, respectively. 
We will see in the next paragraph that the OHC only depends weakly on the mean square displacement $\Delta^2$ that characterizes the thermal disorder in the scattering region and is proportional to the temperature in the equipartition regime \cite{footnote4}. 
Because the OHC is roughly constant in $T$ while the resistivity is linear, the orbital Hall angles for the low and high temperature systems are related to each other as $\Theta_{\rm oH} = \rho \, \sigma_{\rm oH} \sim T$, consistent with previous results for the SHA of Pt \cite{WangL:prl16}. 
The value of the OHC we obtain here is about a factor two lower than other theoretical estimates \cite{Jo:prb18, Salemi:prm22} (but, for what it is worth, actually in better agreement with values extracted from experiment \cite{Sala:prr22}).
Its Fermi energy positions Cr in a deep minimum in $\sigma_{\rm oH}$ calculated as a function of energy (or band-filling).
Below we will see that the form of $\sigma_{\rm oH}(\varepsilon)$ calculated using the scattering formalism tracks the density of states $D(\varepsilon)$ closely and the characteristic minimum occurs for all materials we have studied that have the bcc structure. 
This is in contrast to calculations performed using the Kubo formalism which do not exhibit this minimum in a range of 1 eV around the Fermi energy \cite{Jo:prb18, Salemi:prm22}.

%%%%%%%%%%%%%%%%%%%%%%%%%%%% Fig.4
\begin{figure}[t]
\includegraphics[width=8.6cm]{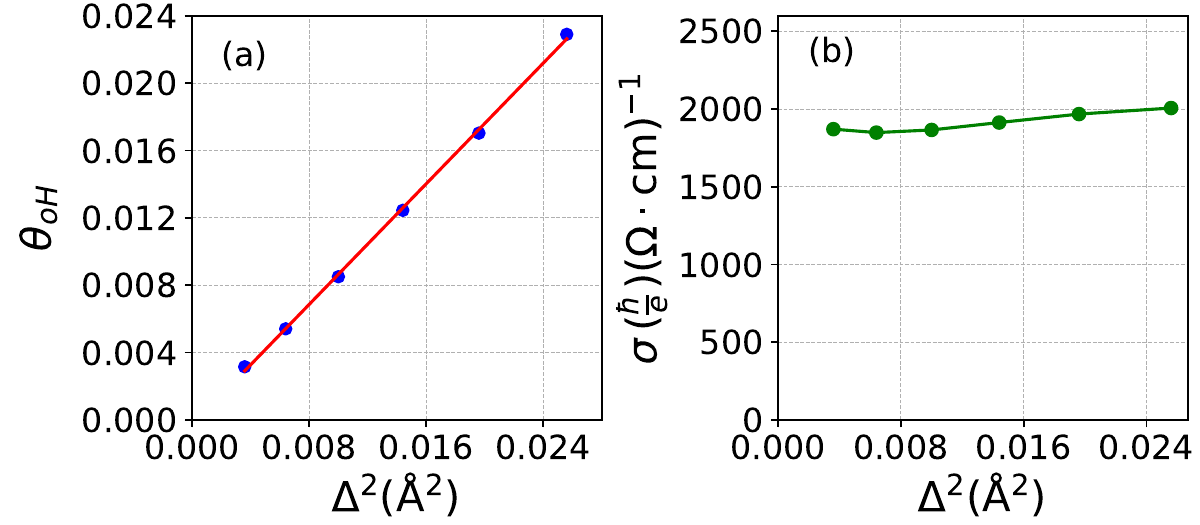}
\caption{The orbital Hall angle (a) and the orbital Hall conductivity (b) in Cr as a function of the mean-square displacement $\Delta^2$ used to model disorder in the scattering region. The mean square displacement is a proxy for temperature.}
\label{fig:OHC_Cr_T_dep}
\end{figure}

\subsubsection{Temperature}

By generating geometries for various values of the root mean square (rms) displacement, we can simulate the effect of thermal lattice disorder on the orbital Hall effect. 
The mechanism generating the transverse currents is intrinsic in the sense that there are no crystal defects or impurities. 
The results shown in \Cref{fig:OHC_Cr_T_dep}(a) show the clear linear dependence of the orbital Hall angle on the mean square displacement of the Cr atoms, which is a proxy for temperature. Because the resistivity is also linear in the mean square displacements, the orbital Hall conductivity is essentially temperature-independent  as shown in \Cref{fig:OHC_Cr_T_dep}(b) \cite{Xiao:prb19}.

\subsection{bcc V}

%%%%%%%%%%%%%%%%%%%%%%%%%%%% Fig.5
\begin{figure}[t]
\includegraphics[width=8.6cm]{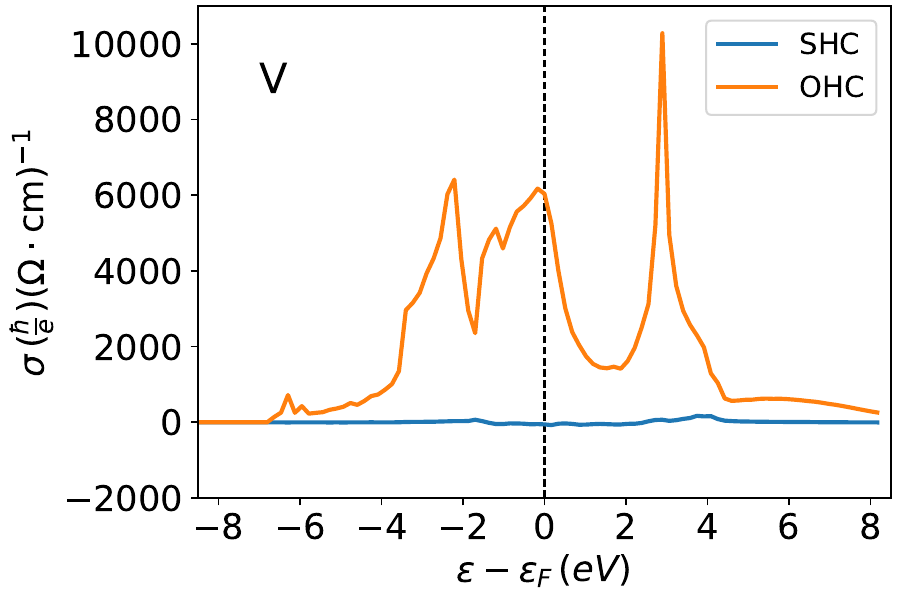}
\caption{Fermi energy dependence of the orbital (spin) Hall conductivity for bulk V at room temperature. The rms displacement of the atoms from their equilibrium position is $\Delta = 0.095 \AA$, corresponding to a resistivity of $\rho = 20.2$ $\mu\Omega \, \text{cm}$.}\label{fig:eF_OHC_V}
\end{figure}

%%%%%%%%%%%%%%%%%%%%%%%%%%%% Fig.6
\begin{figure}[b]
\includegraphics[width=8.6cm]{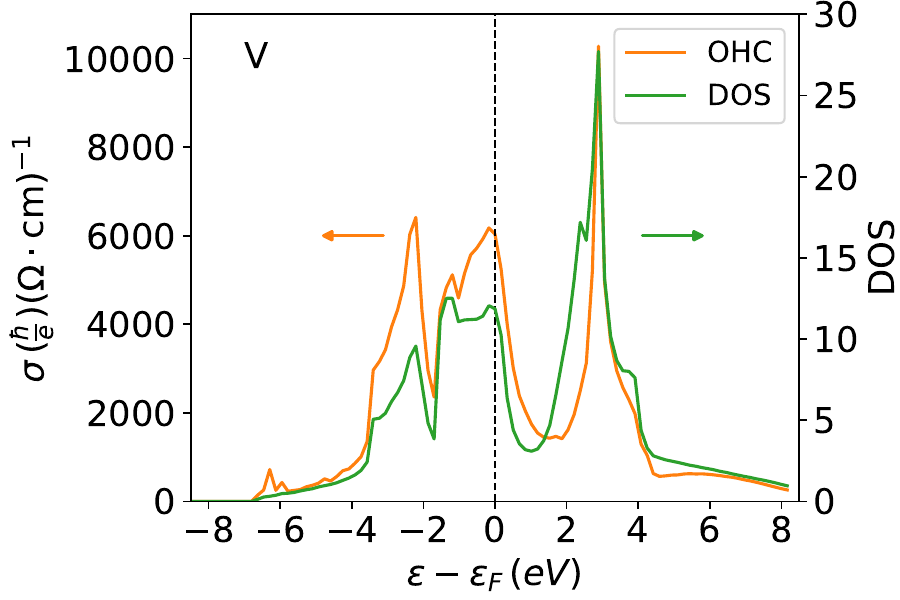}
\caption{Energy dependence of (left-hand axis) the orbital Hall conductivity (OHC) for bulk V together with (right-hand axis) the density of states (DOS), $D(\varepsilon)$.}
\label{fig:eF_OHC_V_wDOS}
\end{figure}

The energy dependence of the OHC for bcc V, shown in \Cref{fig:eF_OHC_V}, is almost identical to that for Cr which is unsurprising because the materials have the same bcc structure and differ by a single electron. 
The larger lattice constant of V does not lead to a narrowing of the 3$d$ band because the 3$d$ orbitals are higher in energy and less localized but otherwise, in a rigid band \cite{Kubler:00} or canonical band \cite{Mackintosh:80, Andersen:85} approximation, this is the only difference. 
One electron less in V lowers the Fermi energy and positions it further up a rapidly increasing $\sigma_{\rm oH}(\varepsilon)$.
This leads to a substantial increase in the OHC of V at its Fermi energy, with $\sigma_{\rm oH} \approx 6 \times 10^3 (\hbar / e) \left(\Omega \, \text{cm}\right)^{-1}$.
The ``canonical'' form of $\sigma_{\rm oH}(\varepsilon)$ for a given crystal structure shows that the value of the OHC can be very sensitive to the band filling. 
In the case of V, shifting the Fermi level up by one eV can result in a reduction of the OHC by a factor three.  
\Cref{fig:eF_OHC_V_wDOS} shows that there is a strong correlation between the OHC and the density of states, a consequence of final state effects in the scattering formalism \cite{Nair:prl21}. 
The value we obtain for the OHC at the Fermi level is comparable to the results reported in other computational studies \cite{Jo:prb18, Salemi:prm22} though the band filling dependence is quite different as already mentioned above.

\subsection{fcc Pt}

\begin{figure}[t]
\centering
\includegraphics[width=8.6cm]{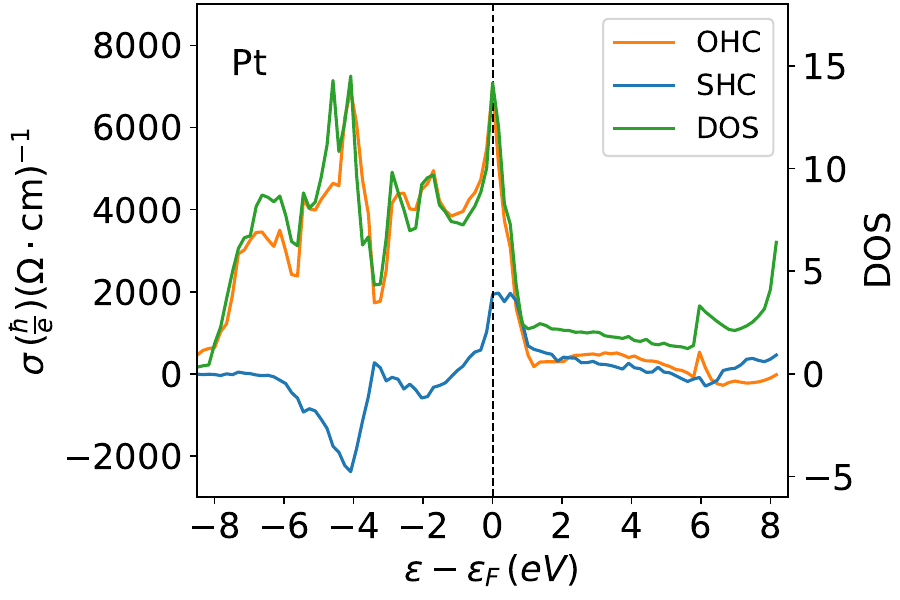}
\caption{Fermi energy dependence of (left-hand axis) the orbital (spin) Hall conductivity for bulk Pt at room temperature, meaning the atoms are randomly displaced from equilibrium by a Gaussian distribution with root mean square value $\Delta = 0.067 \AA$. The DOS is shown with respect to the right-hand axis. 
}
\label{fig:eF_OHC_Pt}
\end{figure}

Our results for the SHC and OHC of fcc Pt are shown in \Cref{fig:eF_OHC_Pt}.
As a function of the band filling, the SHC agrees reasonably well with previous results \cite{Guo:prl08, Jo:prb18, Wesselink:prb19, Salemi:prm22}, with a clear peak at the Fermi level and another, negative, peak centered about 4~eV below the Fermi energy both of which will be slightly reduced when electronic temperature is introduced in the form of the Fermi-Dirac distribution. 
These peak structures can be related to the SOC-induced splitting of orbitally degenerate states at points and along lines of high symmetry \cite{Guo:prl08}. 
In spite of including thermal lattice broadening, we consistently find more structure in both $\sigma_{\rm sH}(\varepsilon)$ and $\sigma_{\rm oH}(\varepsilon)$ with our scattering calculations than found using the Kubo formalism.
Part of this discrepancy may come from the huge value of lifetime broadening (0.4~eV) used by Salemi and Oppeneer \cite{Salemi:prm22} which, while perhaps appropriate for optical experiments, is less obviously justified for transport measurements where the lifetime diverges at the Fermi energy in the absence of disorder; no explanation is given for the failure of $\sigma_{\rm oH}(\varepsilon)$ to vanish for energies below the bottom of the conduction bands. 
The reason for the lack of structure in the $\sigma_{\rm oH}(\varepsilon)$ calculated by Jo {\it et al.} is not clear. 

%%%%%%%%%%%%%%%%%%%%%%%%%%%% Fig.7

Focussing on the value of the OHC at the Fermi level, our scattering calculations predict a larger value of $\sigma_{\rm oH}(\varepsilon_F) \approx 7 \times 10^3 (\hbar / e) (\Omega \, \text{cm})^{-1}$ than the Kubo calculations. 
This large OHC value might partially explain the large variance in experimental values of the SHA in Pt (see Table V in \cite{Wesselink:prb19}). 
For example, in a Pt$|$FM bilayer geometry of Pt adjacent to a ferromagnet (FM), an orbital current generated in the Pt layer is injected simultaneously with a spin current, where the orbital-to-spin conversion at the interface and in the FM layer determines the magnitude of the resulting orbital torque \cite{Go:prr20a, Lee:cmp21, Lee:natc21}.
We expect that the interface conversion of orbital to spin current will depend on details of the interface atomic structure; this should be taken into account in the analysis of spin-orbit torque experiments in any case.

\subsection{hcp Ti}

%%%%%%%%%%%%%%%%%%%%%%%%%%%% Fig.8
\begin{figure}[t]
\centering
\includegraphics[width=8.6cm]{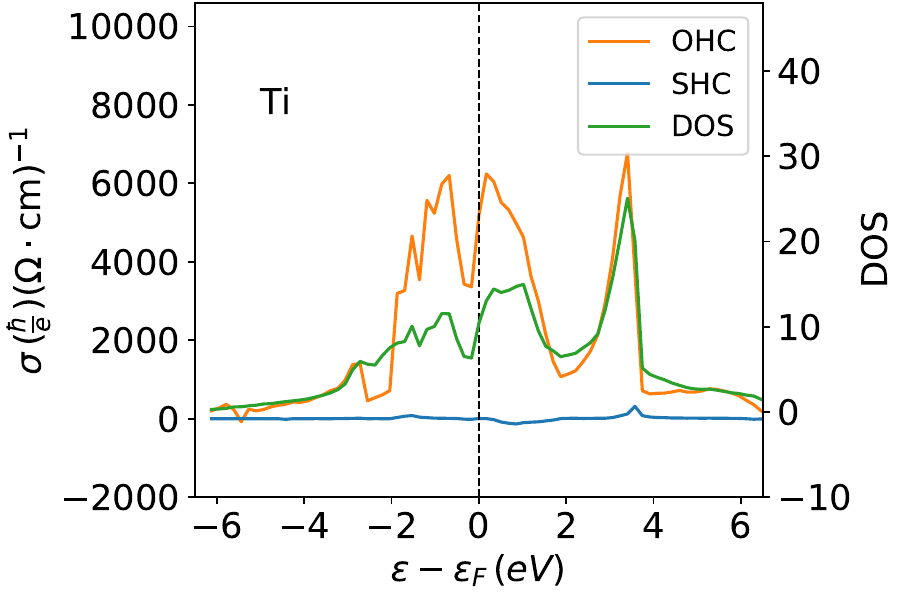}
\caption{Fermi energy dependence of the orbital Hall conductivity for bulk Ti at finite temperature, meaning the atoms are randomly displaced from equilibrium by a Gaussian distribution with root mean square value $\Delta = 0.085 \AA$ ($0.095 \AA$), corresponding to a bulk resistivity of $\rho = 32$ $\mu\Omega$.cm (40.5 $\mu\Omega$.cm).
}
\label{fig:eF_OHC_Ti}
\end{figure}

%%\begin{figure}[t]
%%\includegraphics[width=8.4cm]{/home/max/Work/Papers/NotPublished/OHE_from_transport/v0j/figs/CanonicalDOS.pdf}
%%%PJ\includegraphics[width=8.4cm]{figs/CanonicalDOS.pdf}
%%\caption{Canonical densities of states for the bcc, fcc and hcp structures. }
%%\label{fig:CDOS}
%%\end{figure}

Having performed calculations for transition metal systems with bcc and fcc crystal structures, we complete the set of elementary crystal structures by calculating the OHC for hcp Ti. 
For Cr we explicitly showed that the OHC was independent of temperature. 
We now avoid the (expensive) iterative determination of the rms displacements corresponding to room temperature resistivity for Ti by assuming that the OHC is independent of temperature. 
We model thermal lattice disorder with values of $\Delta = 0.085 \AA$ ($0.095 \AA$), which yield bulk resistivities of $\rho = 32$ $\mu\Omega \,$cm ($40.5 \,\mu\Omega \,$cm). 
These are lower than the room temperature resistivity of $\rho = 48.6$ $\mu \Omega \,$cm \cite{Belskaya:ht05}.
The band filling dependence of the OHC is shown in \Cref{fig:eF_OHC_Ti} and we find (not shown) that its temperature dependence is marginal, once again illustrating the insensitivity to thermal disorder.
The OHC parallels the structure in the DOS as a function of energy but to a lesser extent than for the bcc and fcc structures. 
 The OHC at the Fermi level is $\sigma_{\rm oH} \approx 5 \times 10^3 (\hbar / e) (\Omega \, \text{cm})^{-1}$.

\subsection{fcc Cu}

%%%%%%%%%%%%%%%%%%%%%%%%%%%% Fig.9
\begin{figure}[t]
\centering
\includegraphics[width=8.4cm]{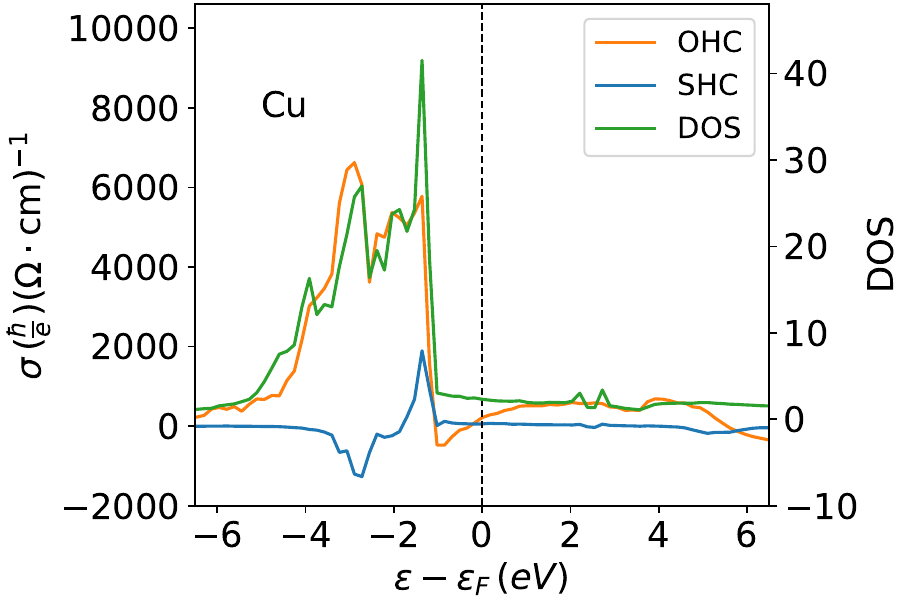}
\caption{Fermi energy dependence of the orbital Hall conductivity for bulk Cu at temperature, meaning the atoms are randomly displaced from equilibrium by a Gaussian distribution with root mean square value $\Delta = 0.08 \AA$.
}
\label{fig:eF_OHC_Cu}
\end{figure}

Repeating the procedure for fcc Cu yields more interesting results, \Cref{fig:eF_OHC_Cu}.
In particular, we see large values of the SHC in the energy region between 1 and 4 eV below the Fermi energy. 
This contradicts the received wisdom that the SOC in $3d$ elements is too weak to yield significant spin Hall conductivities, failing to take into consideration the much smaller bandwidth of $3d$ bands.  
The OHC we find at the Fermi level is $\sigma_{\rm oH}(\varepsilon_F) \approx 200 (\hbar / e) (\Omega \, \text{cm})^{-1}$, much smaller than values of $ \sim 1 \times 10^3 (\hbar / e) (\Omega \, \text{cm})^{-1}$ found using the Kubo formula \cite{Jo:prb18, Salemi:prm22}. 
More recent computational work addressing some of the mathematical problems encountered by the $\mathbf{k}$-space definition of the velocity has even predicted a negative value of $\sigma_{\rm oH}(\varepsilon_F) = -1 \times 10^3 (\hbar / e) (\Omega \, \text{cm})^{-1}$ \cite{Go:prb24}.

\section{Discussion}
We have calculated the orbital Hall conductivity for Ti, V, Cr, Cu and Pt using first-principles scattering theory. 
The expression for the OAM current in terms of flux is inherently interatomic, in contrast to the conventional Kubo-formula expression; we propose that intraatomic currents cannot contribute to the interatomic flux of orbital angular momentum.
 
In this context, the conventional intraatomic OAM operator we used (in the so-called atom-centered approximation) is reasonable especially when using the atomic spheres approximation where all of space is filled with atomic spheres and the volume of the interstitial region is implicitly set to zero, in contrast to FLAPW calculations where a substantial interstitial region is neglected \cite{Salemi:prm22}.
We can attribute differences between results obtained with the two approaches to the explicit exclusion of intraatomic currents in the scattering formalism. 
A direct comparison between the two might yield interesting insights but excluding such contributions from the Kubo formula may be difficult. 
In view of this, there is reasonable, order-of-magnitude agreement for the OHC of 3$d$ metals with the results of Kubo formalism calculations \cite{Jo:prb18, Salemi:prm22}. 
The good agreement between the two approaches for the SHC of Pt is puzzling in the light of the above. 
It may be related to the smallness of the SOC whose effect comes from close to the nucleus whereas orbital quenching  might be expected to depend on details of how interatomic hopping is described; a counterargument is that most mainstream calculations yield very similar band structures.   
We show that the OHC is robust against thermal disorder. Like the SHC, the OHC seems to be more or less independent of temperature corresponding to a linear dependence on temperature of the orbital Hall angle.

The same scattering formulation used here predicts that orbital currents decay on the length scale of the interatomic hopping \cite{Rang:prb24}, in agreement with arguments about ``orbital quenching'' \cite{Tinkham:03, Mohn:03} but not with values of the decay length extracted from experiment.
Therefore, the good agreement between the OHC values we calculate for Cr and values extracted from experiment \cite{Sala:prr22} should probably be taken with a grain of salt until more detailed models are developed to interpret experiment.

%%%
%Additionally, there is the issue of a potential Fermi sea contribution. {\color{red}This can only be understood by insiders ...} The transport calculations are performed with an incoming wave at a certain energy, which is usually the Fermi energy. For the case of the anomalous Hall conductivity, which can be expressed in a comparable fashion to the spin and orbital Hall conductivites, it has been shown that the `non-quantized' contribution is solely determined at the Fermi energy~\cite{Haldane:prl04, WangX:prb07}. Numerical calculations of the same construction has been
%done for the spin Hall conductivity~\cite{Yu:prb21}.
%Though there is a lack of formal proof of the equivalent statement to Haldane's argument for the anomalous Hall effect, we postulate that the same must be true for the orbital Hall effect. Work to show this equivalence is ongoing.
%%%

\subsection{Alternative definition of orbital angular momentum current}
\label{ssec:oam}

In this section, we want to connect with other theoretical work.
This typically starts with a definition of the orbital current operator whereas we arrived at an expression for a current of OAM between two atoms $P$ and $Q$ in \Cref{ss:oc} by applying the continuity equation to the expectation value of $\hat{\ell}$ assuming it to be local, i.e., without invoking a definition.

Our starting point is the expression \eqref{eq:curPQ} for a current of ``X'' between atoms $P$ and $Q$   where, as in \eqref{Eq:K1} and \eqref{Eq:K2}, ``$X$'' is charge ($c$), spin angular momentum ($s$) or OAM ($l$). 
The first term of \eqref{eq:curPQ}, $\left({\bf r}_P - {\bf r}_Q \right) j^{PQ}_X$
\begin{widetext}
\begin{align}
\label{eq:curPQ11}
\left({\bf r}_P - {\bf r}_Q \right) j^{PQ}_X &= \frac{{\bf r}_P - {\bf r}_Q} {i\hbar}
  \Big[\bra{\Psi_P} \hat{X} \hat{H}_{PQ} \ket{\Psi_Q} 
     - \bra{\Psi_Q} \hat{H}_{QP} \hat{X} \ket{\Psi_P} \Big] \nonumber \\
&= \frac{1}{i\hbar} 
\Big[\bra{\Psi_P} {\bf r}_P \hat{X} \hat{H}_{PQ} - \hat{X} \hat{H}_{PQ} {\bf r}_{Q} \ket{\Psi_Q} 
   + \bra{\Psi_Q} {\bf r}_Q \hat{H}_{QP}\hat{X} - \hat{H}_{QP} \hat{X} {\bf r}_P \ket{\Psi_P} \Big] 
\end{align}
can be simplified using the localization of the MTO basis to write $\mathbf{r}_P \ket{\Psi_P} \approx \hat{r} \ket{\Psi_P}$. 
Because the basis functions are not eigenstates of the position operator, the intraatomic position (also called the ``fractional'' part \cite{Turek:prb02}) is lost in this approximation. 
As stated above, we propose that these intraatomic currents do not actually contribute to the spin or orbital Hall currents. Then
%Additionally, the atom-centered operator $X$ is assumed to commute with the atomic position, i.e., $\hat{r} \hat{X} \ket{\Psi_P} \approx r_P \hat{X} \ket{\Psi_P}$ to write
%
\begin{subequations}
\begin{align}
\label{eq:curPQ12}
\left({\bf r}_P - {\bf r}_Q \right) j^{PQ}_X &= \frac{1}{i\hbar} 
 \Big[ \bra{\Psi_P} \hat{r} \hat{X} \hat{H} - \hat{X} \hat{H} \hat{r} \ket{\Psi_Q} 
     + \bra{\Psi_Q} \hat{r} \hat{H} \hat{X} - \hat{H} \hat{X} \hat{r} \ket{\Psi_P} \Big] \\%12a
&= \frac{1}{i\hbar} 
 \Big[ \bra{\Psi_P} \big[ \hat{r}, \hat{X} \hat{H} \big] \ket{\Psi_Q} 
     + \bra{\Psi_Q} \big[ \hat{r}, \hat{H} \hat{X} \big] \ket{\Psi_P} \Big]. %12b
\end{align}
\end{subequations}

\subsubsection{Charge}
\label{ssec:charge}

For a charge current ($X=c$), $\hat{X}= -e \hat{I}$, where $-e$ is the electron charge and $\hat{I}$ is the identity operator. The expression for the charge current becomes
\begin{equation}
\label{eq:lacurPQ}
\left({\bf r}_P - {\bf r}_Q \right) j^{PQ}_c = -\frac{e}{i\hbar} 
\Big[ \bra{\Psi_P} \big[ \hat{r}, \hat{H} \big] \ket{\Psi_Q} 
    + \bra{\Psi_Q} \big[ \hat{r}, \hat{H} \big] \ket{\Psi_P} \Big] 
    = \bra{\Psi_P} \hat{j}_c \ket{\Psi_Q} + \bra{\Psi_Q} \hat{j}_c \ket{\Psi_P}
\end{equation}
\end{widetext}
where we have used the definition of the velocity and current operators
\begin{equation}
\hat{j}_c = -e\hat{v} = -\frac{e}{i\hbar}\big[\hat{r}, \hat{H}\big].
\end{equation}
Since charge conservation requires that $j^{PQ}_c = - j^{QP}_c$, the final expression for $\bar{j}^{PQ}_c$
remains 
\begin{equation}
\bar{j}^{PQ}_c = \bra{\Psi_P} \hat{j}_c \ket{\Psi_Q} + \bra{\Psi_Q} \hat{j}_c \ket{\Psi_P}
\end{equation}
and the description of interatomic electron currents in the transport formulation used here is consistent with the usual definition of the current operator used in the Kubo formalism.

\subsubsection{Spin}
\label{ssec:spin}

For spin, ($X=s$), $\hat{X}=\hat{s}_\alpha = \frac{\hbar}{2} \sigma_\alpha$, where $\sigma_\alpha$ are the Pauli matrices and $\alpha \in \lbrace x,y,z \rbrace$, the spin current between $P$ and $Q$ becomes
\begin{equation}
\label{eq:lacurPQ}
\begin{split}
&\left({\bf r}_P - {\bf r}_Q \right) j^{PQ}_{s \alpha} \\
&= \frac{1}{i\hbar} 
 \Big[ \bra{\Psi_P} \big[ \hat{r}, \hat{s}_\alpha \hat{H} \big] \ket{\Psi_Q} 
     + \bra{\Psi_Q} \big[ \hat{r}, \hat{H} \hat{s}_\alpha \big] \ket{\Psi_P} \Big], %16
\end{split}
\end{equation}
and similarly the spin current between $Q$ and $P$ is
\begin{equation}
\label{eq:lacurPQ2}
\begin{split}
&\left({\bf r}_Q - {\bf r}_P \right) j^{QP}_{s \alpha} \\
&= \frac{1}{i\hbar} 
 \Big[ \bra{\Psi_Q} \big[ \hat{r}, \hat{s}_\alpha \hat{H} \big] \ket{\Psi_P} 
     + \bra{\Psi_P} \big[ \hat{r}, \hat{H} \hat{s}_\alpha \big]\ket{\Psi_Q} \Big], \\ %17
\end{split}
\end{equation}
suggesting the definition of a spin current operator of
\begin{equation}
\begin{split}
\hat{j}_{s \alpha} &= \frac{1}{2i\hbar} \left(\big[ \hat{r}, \hat{s}_\alpha \hat{H} \big] + \big[ \hat{r},\hat{H} \hat{s}_\alpha \big]\right) \\
&= \frac{1}{2i\hbar} \left( \hat{r} \hat{s}_\alpha \hat{H} - \hat{s}_\alpha \hat{H} \hat{r} + \hat{r}\hat{H}\hat{s}_\alpha - \hat{H}\hat{s}_\alpha\hat{r} \right). %18
\end{split}
\end{equation}
This expression is similar, but not identical, to the conventional spin current operator
\begin{equation}
\begin{split}
\hat{j}^{\text{conv}}_{s\alpha} &= \frac{1}{2i\hbar} \left( \hat{s}_\alpha \big[ \hat{r}, \hat{H} \big] + \big[ \hat{r},\hat{H} \big] \hat{s}_\alpha  \right) \\
&= \frac{1}{2i\hbar} \left(\hat{s}_\alpha \hat{r} \hat{H} - \hat{s}_\alpha \hat{H} \hat{r} + \hat{r}\hat{H}\hat{s}_\alpha - \hat{H}\hat{r}\hat{s}_\alpha \right).   %19
\end{split}
\end{equation}
The definitions coincide exactly when $\left[\hat{r},\hat{s}_\alpha\right] = 0$, i.e. when the position and spin operators commute. 
This assumption is automatically satisfied in the approximation that the intraatomic position operator is very small compared to the interatomic position, i.e. $\hat{r}\ket{\Psi_P} \approx {\bf r}_P \ket{\Psi_P}$.
In conclusion, in the scattering formalism, the spin current operator that can be derived from the continuity equation is exactly equivalent to the conventional definition of the spin current operator used elsewhere, notably in the Kubo formalism.

We should mention that there is not yet a consensus about the correct form of the spin current operator. 
A mostly mathematically motivated case exists for the so-called ``proper'' spin current operator \cite{Marcelli:ahp21, Shi:prl06, *ZhangP:prb08}, defined as
\begin{equation}
\hat{j}^{\text{prop}}_{\hat{s}\alpha} = \frac{1}{i\hbar} \big[ \hat{r}\hat{s}_\alpha, \hat{H}\big]
=\frac{1}{i\hbar} \left( \hat{r}\hat{s}_\alpha\hat{H} - \hat{H} \hat{r}\hat{s}_\alpha\right),
\end{equation}
which is only equal to the other two definitions described above when $\big[\hat{H}, \hat{s}_\alpha\big] = 0$, i.e., in the absence of SOC. 
Although this ``proper'' definition has been described as being more accurate than the conventional description, we are not aware of any first-principles calculations of spin Hall conductivities that use it.

\subsubsection{Orbital Angular Momentum}
\label{ssec:oam}

For orbital currents ($X=l$), the operator becomes $\hat{X} = \hat{\ell}_\alpha$, with $x \in \lbrace x,y,z \rbrace$ and the derivation is identical to the spin current case, yielding the orbital current operator
\begin{equation}
\begin{split}
\hat{j}_{l \alpha} &= \frac{1}{2i\hbar} \left(\big[ \hat{r}, \hat{\ell}_\alpha \hat{H} \big] + \big[ \hat{r},\hat{H} \hat{\ell}_\alpha \big]\right) \\
&= \frac{1}{2i\hbar} \left( \hat{r} \hat{\ell}_\alpha \hat{H} - \hat{\ell}_\alpha \hat{H} \hat{r} + \hat{r}\hat{H}\hat{\ell}_\alpha - \hat{H}\hat{\ell}_\alpha\hat{r} \right),
\end{split}
\end{equation}
that is (as in the spin case) similar to, but not identical to, the conventional orbital current operator
\begin{equation}
\begin{split}
\hat{j}^{\text{conv}}_{l\alpha} &= \frac{1}{2i\hbar} \left( \hat{\ell}_\alpha \big[ \hat{r}, \hat{H} \big] + \big[ \hat{r},\hat{H} \big] \hat{\ell}_\alpha  \right) \\
&= \frac{1}{2i\hbar} \left(\hat{\ell}_\alpha \hat{r} \hat{H} - \hat{\ell}_\alpha \hat{H} \hat{r} + \hat{r}\hat{H}\hat{\ell}_\alpha - \hat{H}\hat{r}\hat{\ell}_\alpha \right).
\end{split}
\end{equation}
Again, the expressions are equivalent when $\big[\hat{r},\hat{\ell}_\alpha\big]=0$.
In this picture, the commutation of $\hat{r}$ and $\hat{\ell}$ is mathematically expressed by the approximation of the position operator $\hat{r}\ket{\Psi_P} \approx \mathbf{r}_P \ket{\Psi_P}$, which means that the position operator becomes diagonal in the localized basis, 
together with the assumption that the matrix element $\bra{\Psi_P} \hat{\ell}_\alpha \ket{\Psi_Q}$ is zero for $P\ne Q$.
Only then do $\hat{r}$ and $\hat{\ell}$ commute. 
If $\bra{\Psi_P} \hat{\ell}_\alpha \ket{\Psi_Q}\ne 0$, it becomes impossible to define the OAM current as a flux from $P$ to $Q$, and the current will instead be absorbed into the torque term in the continuity equation.
This is the mathematical equivalent of the statement that it is impossible to define a current of something which is not localized. 
One can only define the flux of something which can be described in local terms. 
Mathematically, one might define a flux from atoms $P$ and $Q$ together, into another set of atoms $R$ and $S$, but the sheer number of fluxes to possible pairs of atoms rapidly becomes intractable.
The method of computing orbital currents presented above is thus limited to systems for which the atomic spheres approximation (ASA) is reasonable, thus excluding materials such as graphene \cite{Bhowal:prb21} where each atom has only three nearest neighbours and, to a lesser extent, the transition metal dichalcogenides \cite{Cysne:prb22} and narrow-gap semiconductors SnTe and PbTe \cite{Pezo:prb22} with six nearest neighbours. 
There should be a clear and unambiguous association of each point in space with a particular atomic center with respect to which the orbital angular momentum can be calculated.

\section{Conclusion}
Starting from the continuity equation, we expressed the orbital Hall effect in terms of interatomic currents of intra-atomic orbital moments and calculated $\sigma_{\rm oH}(\varepsilon)$ for bcc Cr, bcc V, fcc Pt, hcp Ti and fcc Cu, all at or close to room temperature. 
Though the magnitude of the OHC is comparable to values calculated using the Kubo formalism by others \cite{Jo:prb18, Salemi:prm22}, the band filling dependence $\sigma_{\rm oH}(n) \equiv \sigma_{\rm oH}(n(\varepsilon))$ is found to be quite different resulting in an orbital Hall conductivity that is substantially lower for Cr, comparable for Ti, V and Cu and substantially larger for Pt.
$\sigma_{\rm oH}(\varepsilon)$ has much more structure and correlates strongly with the density of states.
The orbital Hall conductivity, like the spin Hall conductivity, is insensitive to changes in temperature, corresponding to linear scaling of both Hall angles and resistivity with temperature. 
That the angular momentum current due to the orbital Hall effect is not suppressed by thermal disorder should be welcome news for technological applications.

\section*{Acknowledgements}
This work was sponsored by NWO Domain Science for the use of supercomputer facilities.

%-------10--------20--------30--------40--------50--------60--------70--------80


\begin{thebibliography}{72}
\bibitem [{foo({\natexlab{a}})}]{footnote2}%
  \BibitemOpen
  \href@noop {} \bibinfo {note} {{For convenience, we
  measure the charge density in units of the electron charge -$e$ (where $e$ is
  a positive quantity), the spin density in units of $\hbar / 2$ and the
  orbital density in units of $\hbar$ so they all become particle densities and
  the fluxes become particle fluxes. In particular, the current and velocity
  operators are identical }}\BibitemShut {NoStop}%
\bibitem [{foo({\natexlab{b}})}]{footnote3}%
  \BibitemOpen
  \href@noop {} \bibinfo {note} {{Assuming that
  $\varepsilon = \varepsilon_{\nu} = \varepsilon_{\rm F}$ and neglecting
  three-center terms}}\BibitemShut {NoStop}%
\bibitem [{foo({\natexlab{c}})}]{footnote4}%
  \BibitemOpen
  \href@noop {} \bibinfo {note} {{Though the onset of the
  equipartition regime is strictly speaking at the Debye temperature
  $\Theta_D$, ``in practice the high-$T$ limit is identified as the
  equipartition regime with linear-in-$T$ resistivity ... but can extend
  practically to about $T>\Theta_D/3$ in Pt, Cu and Au and to about
  $T>\Theta_D/5$ in Al'' \cite{Xiao:prb19}. For typical transition-metal Debye
  temperatures, this means for $T > 100 \,$K \cite{Nair:prl21} }}\BibitemShut
  {NoStop}%
\end{thebibliography}
\end{document}